\UseRawInputEncoding
\documentclass[reprint]{revtex4-2}
\usepackage{amsmath}
\usepackage{caption}
\usepackage{subcaption}
\usepackage{subfig}
\usepackage{graphicx}
\usepackage{epstopdf}
\usepackage{verbatim}
\usepackage{color}
\usepackage{ragged2e}
\usepackage{ulem}
\usepackage{amssymb,amsfonts}
\usepackage{latexsym}
\usepackage{epsfig}
\usepackage{amssymb}
\usepackage{color}
\usepackage{bm}
\usepackage{bbm}
\usepackage{lipsum,mwe}
\usepackage{soul}
\usepackage[thicklines]{cancel}
\newcommand{\be}{\begin{equation}}
\newcommand{\ee}{\end{equation}}

\newcommand{\bea}{\begin{eqnarray}}
\newcommand{\eea}{\end{eqnarray}}

\newcommand{\llangle}{\langle}
\newcommand{\rrangle}{\rangle}

\begin{document}
 
\title{Spin Alignment of Vector Mesons Induced by Local Spin Density Fluctuations}
\author{Kun Xu $^{1,2}$}
\thanks{xukun@bit.edu.cn}

\author{Mei Huang $^{2}$ }
\thanks{huangmei@ucas.ac.cn}

\affiliation{$^{1}$ Beijing Institute of Technology, Beijing 100080, China}
\affiliation{$^{2}$ School of Nuclear Science and Technology, University of Chinese Academy of Sciences, Beijing 100049, China}

\begin{abstract}
We propose a novel mechanism that the spin alignment of vector meson $\phi$ and $K^{*0}$ in most central heavy ion collisions is induced by intrinsic QCD dynamics without external vortical/ magnetic fields. The local spin density fluctuation of quarks due to the axial-vector and tensor interactions between quarks can induce the spin alignments of vector meson. It is found that axial-vector interaction results in $\rho_{00}<1/3$ for the spin alignment of $\phi$ meson, while tensor interaction results in $\rho_{00}>1/3$. We argue that the spin alignment of $K^{*0}$ due to the spin density fluctuation from flavor-mixing would be a signature or hint of instanton.

\end{abstract}
\pacs{Null }
\maketitle

\section{Introduction}
The spin polarization of hyperons and the spin alignment of vector mesons in non-central heavy ion collision were firstly proposed in almost 20 years ago \cite{Liang:2004ph,Liang:2004xn}, and has recently been observed in STAR and ALICE \cite{STAR:2017ckg, ALICE:2019aid,STAR:2022fan}. It was found by STAR that the global spin alignment, denoted by the element of density matrices $\rho_{00}$, of vector meson $\phi$ shows a deviation from $1/3$ \cite{STAR:2022fan} and the deviation increases with the decreasing of the collision energy, while the $\rho_{00}$ of $K^{*0}$ shows non-monotonic behavior, in 20\%-60\% centrality collisions. A small spin alignment of $J/\psi$ was also observed in ALICE \cite{ALICE:2022dyy} for 5.02 TeV Pb-Pb collision.

Understanding such measurements has been a challenge for theorists. Usually the fast rotation and the strong magnetic field created in non-central collisions have been considered as the main causes of spin polarization or alignment. It was found that the global rotation leads to global spin polarization of (anti-)hyperons while magnetic field can induce their splitting \cite{Xu:2022hql}. For spin alignment, it has been found that the global rotation leads to $\rho_{00}<1/3$ \cite{Wei:2023pdf} while the magnetic field induces $\rho_{00}>1/3$ \cite{Sheng:2022ssp}. However, the global rotation or magnetic field in heavy ion collision is believed to be too weak to generate enough $|\rho_{00}-1/3|$ comparable to the experiment data \cite{Yang:2017sdk,Xia:2020tyd}. For example, consider the spin alignment contributed from the global spin polarization $\Lambda/\bar{\Lambda}$, which gives $\rho_{00}\approx \frac{1}{3}-\frac{4}{9}P_{\Lambda}P_{\bar{\Lambda}}$, however, it was found $\frac{4}{9}P_{\Lambda}P_{\bar{\Lambda}} \approx 10^{-6}$, which is too small compared to the experimental data $\sim 10^{-2}$. Recently, Xin-li Sheng proposed that the vector meson field induced by the motion of $s/\bar{s}$ quark can contribute to spin alignment \cite{Sheng:2019kmk,Sheng:2022wsy,Sheng:2023urn}, and the authors of Ref.\cite{Fang:2023bbw} found that the self-energy of quarks can modify the spin polarization and spin alignment. Besides, the global spin alignment has been understood in local axial charge currents fluctuation \cite{Muller:2021hpe}, the helicity polarization\cite{Gao:2021rom},  the local vorticity  and the anisotropic expansion of the fireball\cite{Xia:2020tyd}, the plasma fields \cite{Kumar:2023ghs}, as well as light front quarks\cite{Fu:2023qht}, and so on. The relation between the tensor polarization and quark spin correlations has been explored in \cite{Lv:2024uev}. However, the origin of spin alignment of vector mesons still remains unclear.

On the other hand, spin alignment of vector mesons has also been observed in the most central collisions 0-20\% at STAR \cite{STAR:2022fan},  i.e., $\rho_{00}$ of $\phi$ is smaller than $1/3$ in high energy collision, while larger than $1/3$ at low energy. This observation is confusing from the traditional understanding based on the scenario of rotation and magnetic field, because at high energy and most central collisions, magnetic field, rotation and baryon chemical potential are supposed to be weak, thus $\rho_{00}=1/3$ is expected. This might indicate that the spin alignment of vector mesons is not coming from external rotation and magnetic field.

Therefore, a question arises naturally: Is it possible that the spin alignments of vector mesons are generated by the intrinsic dynamics of QCD? In other words, we want to investigate if the spin alignment could be generated by the inner interaction of quarks, instead of the environment such as rotation or magnetic field. In this work, we propose a new mechanism for the spin alignment of vector meson: the spin density fluctuation of quarks. It is found that the spin of vector mesons depends on the local spin density of quarks but not on the sign of spin polarization, which means that the same spin alignment can be achieved regardless of whether there are more spin-up quarks than spin-down quarks, or vice versa.Thus, we should pay more attention on the local spin polarization instead of the global spin polarization of quarks, because the local fluctuation of spin polarization vanishes after averaging over the system.

Another key ingredient is the interactions between quarks. In principle, the spin structure of a meson consists of contributions from both quarks and gluons. However, it is still quite challenging to consider both quarks and gluons in a meson simultaneously. Thus, we here consider model with quarks only, and the contribution of gluons can be regarded as the coupling constants of quark interactions.  From the view point of four-fermion effective interactions, the scalar interaction is of greatest interest, for its close relation with the spontaneous chiral symmetry breaking in the vacuum and its restoration at finite temperature. However, it is worthy to point out that the axial-vector and tensor interactions are of importance for the study of spin alignment of vector mesons.

This paper is organized as following: in Section II, we show the spin alignment of vector meson related to the spin correlation, and in Section III, we show the axial-vector and tensor operators are appropriate to describe the spin polarization and analyze the axial-vector and tensor interactions induce spin alignment, we show that spin polarization background can be obtained through the spin density fluctuation induced by axial-vector interaction, while it cannot be obtained through a mean field approximation. In Section IV, we argue two possible origination of the axial-vector and tensor interaction, one-Gluon-Exchange process and (anti-)instanton, and the latter one is favored because of the flavor-mixing results.

\section{Global spin polarization under Mean Field Approximation}
From the view of simple quark model, the spin of vector meson depends on the spins of its constituent quarks. And in the heavy ion collision, QGP is generated and as the fireball expands, its temperature decreases and mesons are emitted at freeze-out temperature from a local equilibrium region, the spin density, i.e., the density of spin-up minus the density of spin-down, of quark and anti-quark determines the distribution of emitted mesons with different spins. Thus, we firstly need to find appropriate quantities to describe the spin density. In the following,  we will show that the  axial-vector operator is related to the spin polarization(SP) while tensor operator to the magnetic moment polarization(MMP), both are crucial for describing the spin alignment of vector mesons.

\subsection{Axial-Vector and Tensor operator}

Consider the axial-vector operator $\bar{\psi}\gamma^{\mu}\gamma^5\psi$ for free quark of one flavor, here we  take the $z$-axis without loss of generality, i.e., $\mu=3$.  With the standard procedure of quantum field theory  and use the form of quark field expanded in momentum space, we could easily find that the integral of axial-vector operator over all space could be divided into four terms:
\begin{equation}
    \int_x \bar{\psi}(x) \gamma^3\gamma^5 \psi(x) = J^{\uparrow}_{+}-J^{\downarrow}_{+}+J^{\uparrow}_{-}-J^{\downarrow}_{-}=:\delta J, 
\end{equation}
and here we show $J^{\uparrow}_{+}$ explicitly as an example:
\begin{eqnarray}
J^{\uparrow}_{+}&=&N_c\int_p \frac{Ep_{z}^2+m_qp_{\perp}^2}{E\mathbf{p}^2}a_p^{\uparrow\dagger} a_p^{\uparrow},
\end{eqnarray}
where $N_c=3$ is number of color and  we define the abbreviation: 
\begin{equation}
    \int_p=\int \frac{d^3p}{(2\pi)^3},\quad \int_x=\int d^3x,
\end{equation} 
and in the small momentum limit, i.e., $\mathbf{p} \rightarrow 0$, or large mass limit, $J^{\uparrow}_{+}=N_c\int_p a_p^{\uparrow\dagger} a_p^{\uparrow}$ is the total number of quark with spin-up, and the meaning of $\delta J$ is clear: the net-spin angular momentum(up to a factor $\hbar/2$). And at medium or large momentum, $J^{\uparrow}_{+}$ is not exact the number of net-spins,  however, we can still relate it to the number of net-spins:
\begin{equation}
    J^{\uparrow/\downarrow}_{+/-}=c N^{\uparrow/\downarrow}_{+/-},
\end{equation}
where, for example, the number of net-spins for quark is:
\begin{equation}
    N^{\uparrow/\downarrow}_{+}=N_c\int_p a_p^{\uparrow\dagger} a_p^{\uparrow},
\end{equation}
and $c$ is a function of temperature and quark mass, and in a equilibrium system, for example, $c\approx 0.22$ for $T=0.15$GeV and $m_q=0.33$GeV. Thus, $\delta J$ can be regarded as spin polarization.

Similarly, the tensor operator $\Bar{\psi}\Sigma^3\psi=\Bar{\psi}i\gamma^1\gamma^2\psi$ indicates the imbalance between magnetic-moment-up and -down, i.e., magnetic moment polarization\sout{(MMP)}, instead of spin polarization in the case of axial-vector operator, which can be seen from:
\begin{equation}
    \int_x \bar{\psi}(x) \Sigma^3 \psi(x) = M^{\uparrow}_{+}-M^{\downarrow}_{+}-(M^{\uparrow}_{-}-M^{\downarrow}_{-})=:\delta M, 
\end{equation}
and here we also show $M^{\uparrow}_{+}$ explicitly as an example:
\begin{eqnarray}
M^{\uparrow}_{+}&=&N_c\int_p \frac{Ep_{\perp}^2+m_qp_{z}^2}{E\mathbf{p}^2}a_p^{\uparrow\dagger} a_p^{\uparrow}
\end{eqnarray}
and similarly, use $M^{\uparrow/\downarrow}_{+/-}=d N^{\uparrow/\downarrow}_{+/-}$, $d\approx 0.29$ for $T=0.15$GeV and $m_q=0.33$GeV.

\subsection{Global Quark Spin Polarization}
The simplest case for spin polarization is a global and homogeneous spin polarization, and it is straightforward to consider that the SP is induced spontaneously from the quark interactions, where a global and homogeneous background, i.e., condensate, can be obtained. Here we assume a NJL-type four-fermion interaction of axial-vector channel, $G_A ( \Bar{\psi}\gamma^{\mu}\gamma^5 \psi )^2$, where $\mu=3$ is taken in the following. All the following argument can be applied to the case of tensor interaction, $G_T ( \Bar{\psi}\Sigma^3 \psi )^2$, and we neglect this case here. Then the Lagrangian for quarks with NJL-type scalar and axial-vector interactions has form of \cite{Klevansky:1992qe,Hatsuda:1994pi}:
\begin{equation}
    \mathcal{L}=\Bar{\psi} (i\partial\!\!\! /-m_0) \psi +G_S ( \Bar{\psi}\psi )^2+G_A (\Bar{\psi}\gamma^3\gamma^5\psi )^2,
\end{equation}
where $m_0$ is current quark mass. Similar to the definition of chiral condensate $\langle \Bar{\psi}\psi\rangle$, we here also define the axial-vector condensate $\langle \Bar{\psi}\gamma^3\gamma^5 \psi \rangle$, then we can write down the new Lagrangian under mean field approximation(MFA) as usually used in NJL model:
\begin{eqnarray}
   \mathcal{L} =  \Bar{\psi} (i\partial \!\!\! /-m_q + a \gamma^3\gamma^5) \psi - G_S \langle \Bar{\psi}\psi\rangle^2 - G_A\langle \Bar{\psi}\gamma^3\gamma^5 \psi \rangle^2, \nonumber\\
\end{eqnarray}
where we defined the effective quark mass and SP potential as:
\begin{equation}
    m_q=m_0-2G_S\langle \Bar{\psi}\psi\rangle,\quad a=-2G_A\langle \Bar{\psi}\gamma^3\gamma^5 \psi \rangle,
\end{equation}
consider a quark system at finite temperature $T$ and volume $V$, and the grand canonical ensemble is usually used to describe the fireball in heavy ion collision\cite{Yagi:2005yb}, and more details on finite temperature field theory can be found in the textbook\cite{Kapusta:2006pm}. Then the thermodynamic potential
\begin{equation}
    \Omega=-\frac{T}{V}\ln Z,
\end{equation}
is obtained, where $Z$ is the partition function at finite temperature. Then the SP potential $a$ as well as the effective mass $m_q$ can be obtained by solving the gap equations:
\begin{equation}
    \frac{\partial \Omega}{\partial \langle \Bar{\psi}\psi\rangle}=0, \quad \frac{\partial \Omega}{\partial \langle \Bar{\psi}\gamma^3\gamma^5 \psi \rangle}=0.
\end{equation}

Usually, the chiral condensate is not zero at low temperature, $\langle \Bar{\psi}\psi \rangle \neq 0$, similarly, as discussed above, a non-zero tensor condensate, i.e.,  $\langle \Bar{\psi}\gamma^3\gamma^5 \psi \rangle \neq 0$,  is expected. However, it is found numerically that for $G_A\lesssim 2G_S$,  $\langle \Bar{\psi}\gamma^3\gamma^5 \psi \rangle$ is always zero at finite temperature. And for $G_A\gtrsim 2G_S$, $\langle \Bar{\psi}\gamma^3\gamma^5 \psi \rangle$ could be  non-zero, however, $\langle \Bar{\psi}\gamma^3\gamma^5 \psi \rangle$ and $\langle \Bar{\psi}\psi\rangle$ cannot be non-zero simultaneously, in this case where $m_q=0$ and $a\neq 0$, only spin-up of quarks exists, which leads to $\rho_{00} \sim 1$ of vector mesons. 

\begin{figure}
 \centering
	\includegraphics[width=0.5\textwidth]{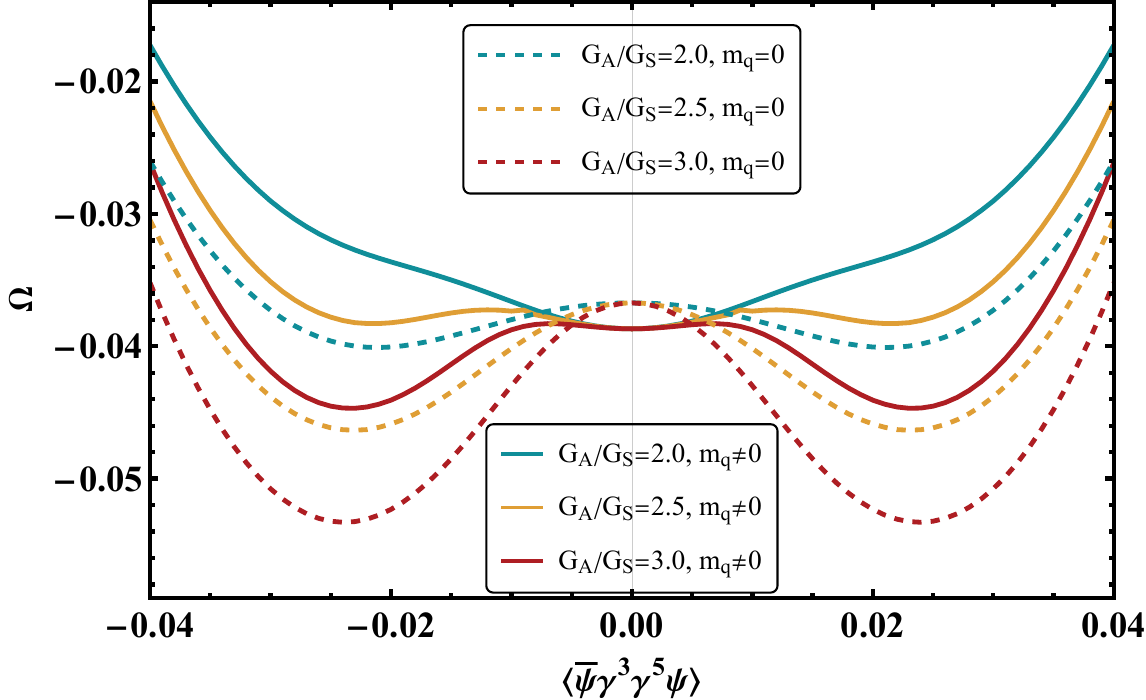}
	\caption{\justifying Thermodynamic potential $\Omega$ with different axial-vector interaction coupling constant and quark effective mass, as a function of axial-vector condensate. This is at zero temperature.}
\label{fig:Omega}
\end{figure}

We show the case of zero temperature and zero baryon chemical potential in FIG.\ref{fig:Omega}, where thermodynamic potential as a function of  $\langle \Bar{\psi}\gamma^3\gamma^5 \psi \rangle$, with different axial-vector interaction coupling constants and quark effective mass, are plotted. For $G_A=2G_S$, as plotted by green solid and dashed lines, the minimal of $\Omega$ lie in $m_{l/s} \neq 0$ and $\langle \Bar{\psi}\gamma^3\gamma^5 \psi \rangle=0$. And for  $G_A=2.5G_S$ or larger, the minimal of $\Omega$ lie in $m_{l/s} = 0$ and $\langle \Bar{\psi}\gamma^3\gamma^5 \psi \rangle \neq 0$, as the yellow/red solid or dashed lines shown. Similar study can also be found in Ref.\cite{Maruyama:2017mqv}. Considering that the axial-vector interaction between quarks is supposed to be week, i.e., $G_A \ll G_s$, $\langle \Bar{\psi}\gamma^3\gamma^5 \psi \rangle=0$ thus no global SP mean field could be obtained at finite temperature. The same conclusion for tensor interaction and $\langle \Bar{\psi}i\gamma^1\gamma^2 \psi \rangle$ can also be obtained.

\section{spin alignment induced by quark spin polarization}
\subsection{spin correlation and spin fluctuation of quarks}
Let's consider how the imbalance between spin-ups and spin-downs of quarks induce the spin alignment of vector meson, and here we ignore the effect from orbital angular momentum. In heavy ion collision, the $\phi$ meson are generated when the temperature decreases and reach the freeze-out condition. For $\phi$ meson with $s_z=1$ and $s_z=-1$, the possibilities can be expressed as:
\begin{equation}
    P_{1}=\frac{N^{\uparrow}_s N^{\uparrow}_{\bar{s}}}{N_s N_{\bar{s}}},~~ P_{-1}=\frac{N^{\downarrow}_s N^{\downarrow}_{\bar{s}}}{N_s N_{\bar{s}}},
\end{equation}
where $N^{\uparrow/\downarrow}_{s/\bar{s}}$ is the number of spin-up/-down $s/\bar{s}$ quark in given volume, and $N_{s/\bar{s}}=N^{\uparrow}_{s/\bar{s}}+N^{\downarrow}_{s/\bar{s}}$ is the total number of $s/\bar{s}$ quark. 
And the possibility for $s_z=0$ is similar:
\begin{equation}
     P_0=\frac{1}{2}\frac{N^{\uparrow}_s N^{\downarrow}_{\bar{s}}+N^{\downarrow}_s N^{\uparrow}_{\bar{s}}}{N_s N_{\bar{s}}},
\end{equation}
because that there are two states for $s_z=0$. Then the spin alignment $\rho_{00}$ can be simply expressed as the possibility to find a $\phi(s_z=0)$ from all $\phi$s:
\begin{equation}
    \rho_{00}=\frac{P_0}{P_{1}+P_{-1}+P_0}.
\end{equation}
In a equilibrium and non-polarized system, $\rho_{00}=1/3$ is obvious. However, due to the global spin polarization or spin fluctuation or any other possible reasons, $N^{\uparrow}_{s/\bar{s}} \neq N^{\downarrow}_{s/\bar{s}}$ thus $\rho_{00} \neq 1/3$ could happen. Define $N_s=N^{\uparrow}_s+N^{\downarrow}_s$, which is the total density of $s$ quark with both spin-up and spin-down, and $\delta N_s=N^{\uparrow}_s-N^{\downarrow}_s$, which is the number of net-spin of $s$ quark. Assume that $\delta N_s$ is small, i.e., $\delta N_s/N_s \ll 1$, then we obtain:
\begin{equation}
\label{eq:rho00}
    \rho_{00}\approx \frac{1}{3}-\frac{4}{9}\frac{\delta N_s}{N_s}\frac{\delta N_{\bar{s}}}{N_{\bar{s}}},
\end{equation}
it is clear that the spin polarization of $s/\bar{s}$ quark directly lead to spin alignment of $\phi$ meson, and the deviation of $\rho_{00}$ from $1/3$, i.e., $\delta \rho_{00}=\rho_{00}-1/3$, could either be positive or negative, which depends on the correlation of net-spin of $s$ quark and $\bar{s}$ quark. Similar formula also can be found in \cite{Lv:2024uev}. In a local small region, for example,  if spin of $s$ quark tends to be same with $\bar{s}$, let's say induced by global rotation, then  $\llangle \delta N_s  \delta N_{\bar{s}} \rrangle >0$, thus  $\delta \rho_{00}<1/3$; If magnetic field is presented, then  $\llangle \delta N_s  \delta N_{\bar{s}} \rrangle <0$ is obtained because that $s$ and $\bar{s}$ has opposite charge, thus  $\delta \rho_{00}>1/3$. 

It is worthy to point out that here we don't consider the difference between $s$ and $\bar{s}$, i.e., at $\mu_B=0$. And we have found that non-zero baryon chemical potential leads to larger deviation of $\rho_{00}$ from $1/3$, however, here we will not make further discussion. Then consider another vector meson $K^{*0}$, which is composed of $d$ and $\bar{s}$, and its spin alignment has a similar form: 
\begin{equation}
    \rho_{00}\approx \frac{1}{3}-\frac{4}{9}\frac{\delta N_d}{N_d}\frac{\delta N_{\bar{s}}}{N_{\bar{s}}},
\end{equation}
now it is the correlation between the net-spin of $d$ quark and $\bar{s}$ quark that affect the spin alignment of $K^{*0}$.

\begin{figure}
 \centering
	\includegraphics[width=0.5\textwidth]{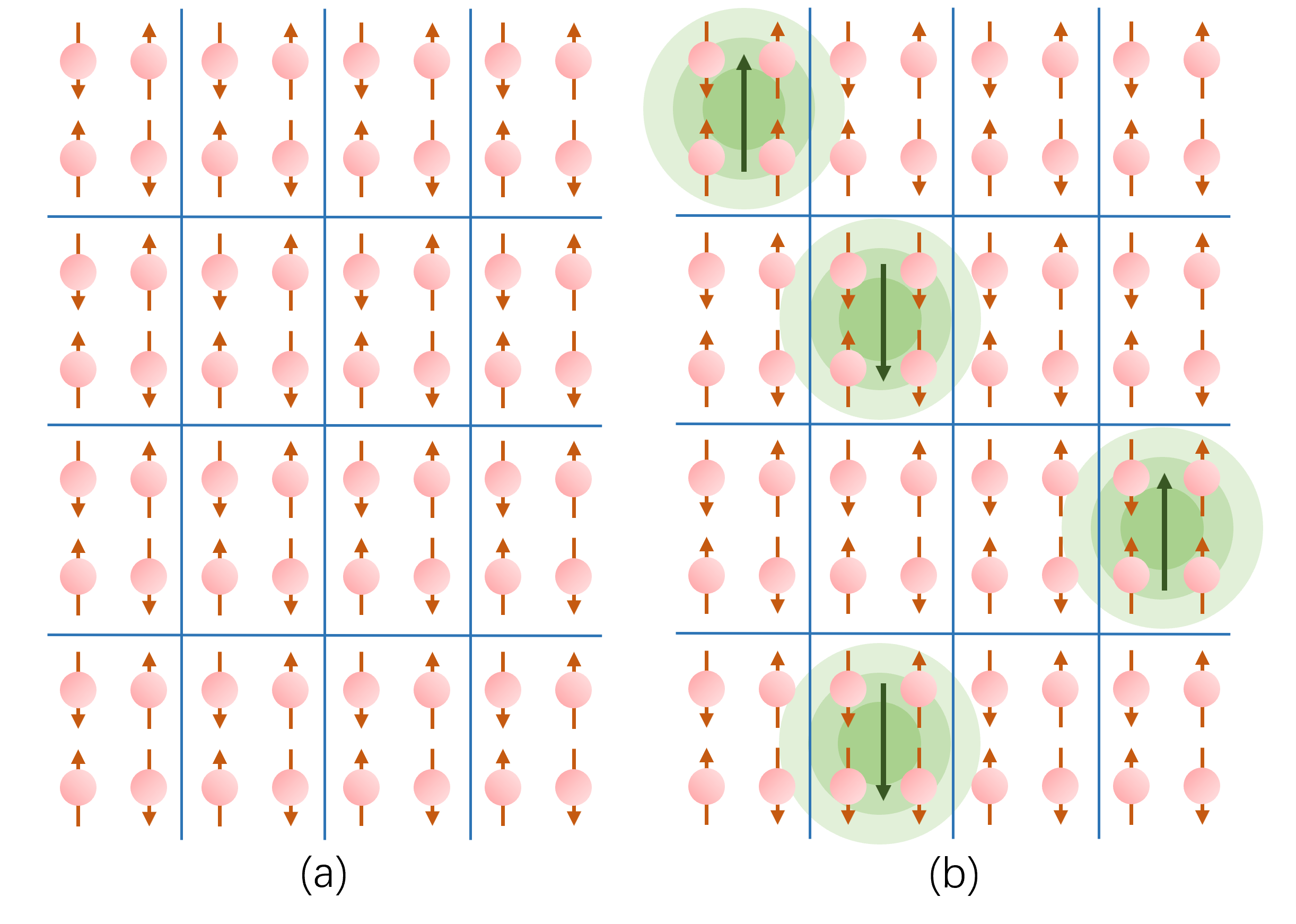}
	\caption{\justifying  Schematic diagram of quark spin fluctuation. (a) No global and local spin polarization. However, the local spin polarization of quark could be non-zero for a quantum system at finite temperature due to the fluctuation even no global spin polarization, as shown in (b).}
\label{fig:fluctuation}
\end{figure}
Although non-zero $\llangle \Bar{\psi}\gamma^{3}\gamma^{5}\psi \rrangle$ can not be obtained spontaneously with the axial-vector interaction, as we have proved in previous section, $\Bar{\psi}\gamma^{3}\gamma^{5}\psi$ is still not exactly zero at any local region of the system due to the fluctuation nature of a quantum system at finite temperature, as illustrated in FIG.\ref{fig:fluctuation}. On the other hand,  the spin alignment $\delta\rho_{00}$ doesn't require a non-zero global and homogeneous $\llangle \Bar{\psi}\gamma^{3}\gamma^{5}\psi \rrangle$ but the spin correlation $\llangle \delta N_s \delta N_{\bar{s}} \rrangle \neq 0$. Consider the free quark gas where the interaction between quarks is ignored, then the fluctuation of quark and anti-quark is completely independent, i.e., $\llangle \delta N_s \delta N_{\bar{s}} \rrangle = \llangle \delta N_s  \rrangle \llangle \delta N_{\bar{s}} \rrangle=0$, thus no spin alignment of vector meson will appear. If we turn on the quark interaction,  let's say a attraction between spins, then the spin of quark as well as anti-quark tend to point to the same direction in a local region, i.e., $\llangle \delta N_s \delta N_{\bar{s}} \rrangle 
>0$, which means the sign of local spin polarization, $\delta N$, doesn't affect the sign of spin alignment, $\delta\rho_{00}$, and the contribution from local spin polarization will not be washed  out by averaging over all events or the entire system.

\subsection{Spin Alignment of $\phi$}
Because of the direct relation between spin alignment of spin correlation as shown in Eq.(\ref{eq:rho00}), now we show how to calculate $\llangle \delta N_s \delta N_{\bar{s}} \rrangle$. Firstly, use the definition of axial-vector and tensor operators, we found the following equations:
\begin{widetext}
\begin{eqnarray}
   \llangle \delta J^2 \rrangle = \llangle  (\delta J_s+\delta J_{\bar{s}})^2 \rrangle
    =\llangle \delta J_s^2  \rrangle+\llangle \delta J_{\bar{s}}^2  \rrangle+2\llangle \delta J_s \delta J_{\bar{s}}  \rrangle 
    =c^2\biggl(\llangle \delta N_s^2  \rrangle+\llangle \delta N_{\bar{s}}^2  \rrangle+2\llangle \delta N_s \delta N_{\bar{s}} \rrangle \biggr),
\end{eqnarray}
similarly for tensor operator:
\begin{eqnarray}
  \llangle \delta M^2\rrangle = \llangle (\delta M_s-\delta M_{\bar{s}})^2  \rrangle 
    =\llangle \delta M_s^2  \rrangle+\llangle \delta M_{\bar{s}}^2  \rrangle-2\llangle \delta M_s \delta M_{\bar{s}}  \rrangle
    =d^2 \biggl(\llangle \delta N_s^2  \rrangle+\llangle \delta N_{\bar{s}}^2  \rrangle-2\llangle \delta N_s \delta N_{\bar{s}} \rrangle \biggr),
\end{eqnarray}
\end{widetext}
thus the spin correlation can be obtained:
\begin{equation}
   4 \llangle \delta N_{s}\delta N_{\bar{s}}  \rrangle= \frac{1}{c^2}\llangle \delta J^2 \rrangle-  \frac{1}{d^2}\llangle \delta M^2 \rrangle,
\end{equation}
which relates to the axial-vector and the tensor operators, and that's why we consider these two operators. Now, we show how to calculate $\llangle \delta J^2 \rrangle$, and the process is total same for $\llangle \delta M^2 \rrangle$. Consider quarks in a spin polarization background where a non-zero spin potential $a$ is presented, then define the  thermodynamic potential:
\begin{equation}
    \Omega=-\frac{T}{V}\text{ln}Z(a),
\end{equation}
where $Z(a)$ is the partition function at finite temperature with effective quark mass $m$ and the effective spin potential $a$ as well as the axial-vector interaction $G_A (\Bar{\psi}\gamma^3\gamma^5\psi )^2$, which has form of
\begin{eqnarray}
    Z(a)&=&\int D\bar{\psi}D\psi \;\text{exp} \biggl \{\int d^3x d\tau \Bar{\psi} (i\partial\!\!\! /-m+a\gamma^3\gamma^5) \psi \nonumber\\
    &&+G_A (\Bar{\psi}\gamma^3\gamma^5\psi )^2 \biggr \},
\end{eqnarray}
where $\tau=it$ is imaginary time, and the integral over $\tau$ is from 0 to $\beta$, where $\beta=1/T$ \cite{Kapusta:2006pm}. And now the partition function is a function of $a$. Then we consider the second derivative of $\Omega$ with respect to $a$, which is found to have two terms:
\begin{equation}
    \mathcal{I}=\frac{\partial^2 \Omega}{\partial a^2}=\mathcal{I}_1+\mathcal{I}_2,
\end{equation}
where
\begin{equation}
    \mathcal{I}_1=\frac{T}{V}\biggl(\int d^3x d\tau \llangle \bar{\psi} \gamma^3\gamma^5\psi \rrangle \biggr)^2=\frac{1}{TV}  \llangle \delta J\rrangle^2,
\end{equation}
which is zero for $a=0$, and the second term:
\begin{eqnarray}
\mathcal{I}_2 &=&-\frac{T}{V}\int d^3x d\tau d^3x' d\tau' \llangle (\bar{\psi} \gamma^3\gamma^5\psi)_{\tau,x}(\bar{\psi} \gamma^3\gamma^5\psi)_{\tau',x'}  \rrangle \nonumber\\
&=&-\frac{1}{TV} \llangle \delta J^2  \rrangle,
\end{eqnarray}
If we take $a\rightarrow 0$, then the first term will be zero because $\lim_{a\rightarrow 0} \llangle N_s \rrangle = 0$, while the second term gives non-zero contribution, which is exactly what we want:
\begin{equation}
    \label{eq:deltaJ2}
   \llangle  (\delta J)^2 \rrangle=-VT \frac{\partial^2 }{\partial a^2}\Omega(a)|_{a\rightarrow 0}.
\end{equation}
Actually, above calculation is nothing but:
\begin{eqnarray}
    \llangle  (\delta J-\llangle  \delta J\rrangle)^2 \rrangle&=& \llangle  \delta J^2 \rrangle + \llangle  \llangle\delta J\rrangle^2 \rrangle-2\llangle  \delta J \llangle\delta J\rrangle\rrangle \nonumber\\
    &\overset{a\rightarrow 0}{=}& \llangle  \delta J^2 \rrangle .
\end{eqnarray}

Then consider the thermodynamic potential, the Feynman diagram of which is shown in FIG.\ref{fig:omegaFD} and only the leading order of axial-vector interaction with coupling constant $G_A$ is considered. The first one is of order $N_c^2 G_{A}$ while the second one $N_c G_{A}$. And similar with the case of chiral condensate, we only consider the first Feynman diagram, which is of form:
\begin{figure}
 \centering
	\includegraphics[width=0.5\textwidth]{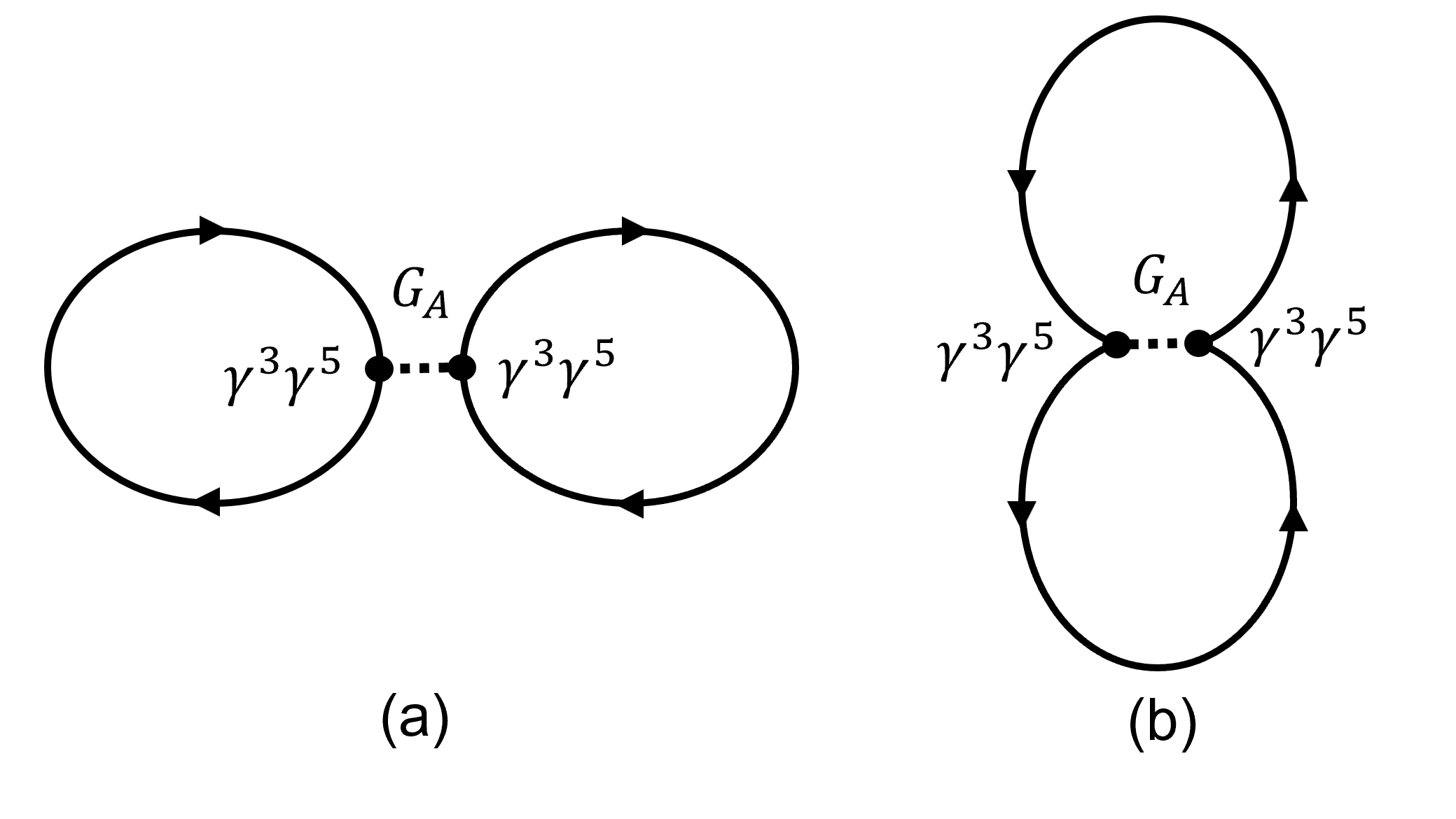}
	\caption{\justifying The thermodynamic potential in the leading order with respect to the axial-vector interaction. Diagram (a) is of order $G_AN_c^2$ while (b) is $G_A N_c$.}
\label{fig:omegaFD}
\end{figure}
\begin{equation}
    \Omega= - G_A N_c^2 \Tilde{L}^2,
\end{equation}
where 
\begin{equation}
    \Tilde{L}= T\sum_{n}\int \frac{d^3p}{(2\pi)^3} \text{Tr}[S(a)\gamma^3\gamma^5],
\end{equation}
and $\text{Tr}$ is trace over Dirac space. The propagator of quark with spin potential:
\begin{equation}
    S(a)=\frac{i}{p\!\!\!/-m+a\gamma^3\gamma^5},
\end{equation}
with $p_0=i\omega_n$ and $\omega_n=(2n+1)\pi T$. Considering that $a \rightarrow 0$ will be taken at last, we take this limit first:
\begin{eqnarray}
   \lim_{a\rightarrow 0}  \Tilde{L}&=& -4aT\sum_{n}\int \frac{d^3p}{(2\pi)^3} \frac{p_0^2-p_{\perp}^2+p_3^2+m^2}{(p^2-M^2)^2}\nonumber\\
    &=&-4a\int \frac{d^3p}{(2\pi)^3}\frac{p_{\perp}^2}{2E^3}2f(E):=4a L,
\end{eqnarray}
where $f(E)=1/(\text{e}^{E/T}+1)$ is Fermi-Dirac distribution, $p_{\perp}$ is the momentum in the $x-y$ plane while $E=\sqrt{p_{\perp}^2+p_3^2+m^2}$ is energy. Now there is no $a$-independent term, and this is reasonable because that there is of course no spin polarization at zero spin potential $a=0$. Now we obtain the  thermodynamic potential:
\begin{eqnarray}
   \lim_{a\rightarrow 0}\Omega = -16N_c^2G_A L^2 a^2,
\end{eqnarray}
together with Eq.(\ref{eq:deltaJ2}), we finally obtain:
\begin{eqnarray}
    \llangle \delta J^2 \rrangle=32N_c^2G_A VTL^2.
\end{eqnarray}
The process for calculating  $\llangle (\delta M)^2  \rrangle$ is same, however, notice:
\begin{eqnarray}
\label{eq:mmpeq}
   T\sum_{n}\int  \frac{d^3p}{(2\pi)^3}\text{Tr}[(S'(\kappa)\gamma^1\gamma^2)]=0,
\end{eqnarray}
where 
\begin{equation}
    S'(\kappa)=\frac{1}{p\!\!\!/-m+\kappa i\gamma^1\gamma^2},
\end{equation}
is propagator of quark in a magnetic moment polarization background with $\kappa$ the MMP potential, and Eq.(\ref{eq:mmpeq}) indicates the axial-vector interaction can not induce local magnetic moment polarization thus $\llangle (\delta M)^2  \rrangle = 0$. As a result, the spin alignment of $\phi$ induced by the spin fluctuation with axial-vector is:
\begin{eqnarray}
    \delta\rho_{00}(\phi) &\approx& -\frac{4}{9}\frac{\llangle \delta N_{s}\delta N_{\bar{s}}  \rrangle}{ N_s N_{\bar{s}}} 
    =-\frac{32}{9c^2}\frac{T}{V}\frac{N_c^2G_A}{\rho_s^2}  L^2,
\end{eqnarray}
which is negative and $\rho_s=\rho_{\bar{s}}=\int_p f(p)$ is the number density of $s/\bar{s}$ quark  in a equilibrium system.

On the other hand, if we consider the the tensor interaction $G_T(\bar{\psi} i\gamma^1\gamma^2\psi)^2$ instead of axial-vector interaction, following above process we found:
\begin{equation}
     \llangle \delta J^2 \rrangle = 0, \llangle \delta M^2 \rrangle \neq 0,
\end{equation}
thus the spin alignment induced by tensor interaction is positive:
\begin{equation}
    \delta\rho_{00}(\phi) \approx \frac{32}{9d^2}\frac{T}{V}\frac{N_c^2G_T}{\rho_s^2}  L'^2,
\end{equation}
where
\begin{equation}
    L'=\int \frac{d^3p}{(2\pi)^3}\frac{p_{3}^2}{2E^3}2f(E).
\end{equation}
In a nutshell, by the local spin fluctuation, the spin alignment of $\phi$ can be induced by the axial-vector interaction leading to $\rho_{00}<1/3$ while tensor interaction leading to $\rho_{00}>1/3$.

\subsection{Spin alignment of $K^{*0}$ }
$K^{*0}$ meson is similar to $\phi$, and the only difference is replacing the composed $s$ quark by $d$ quark, and the spin correlation of $s$ and $\bar{d}$ is of form:
\begin{eqnarray}
   4 \llangle \delta N_{d}\delta N_{\bar{s}}  \rrangle &=& \frac{1}{c_d c_s}\llangle \delta J_d \delta J_s \rrangle -\frac{1}{d_dd_s}\llangle \delta M_d \delta M_s\rrangle,
\end{eqnarray}
where we have simply set $\llangle \delta N_{d}\delta N_{\bar{s}} \rrangle=\llangle \delta N_{s}\delta N_{\bar{d}} \rrangle$. For a non-flavor-mixing-interacting system, the spin polarization of  $s$ quark and $d$ quark could be completely independent, and it is obvious that no spin alignment of $K^{*0}$ could arise, even the local spin density for $s$ or $d$ quark is  non-zero. In other words, the observed spin alignment of $K^{*0}$ requires flavor-mixing axial-vector interaction in this framework:
\begin{equation}
\label{equ:flavormixing}
   G_{A}^{mix}(\bar{s} \gamma^{3}\gamma^5 s)(\bar{d} \gamma_{3}\gamma^5 d),
\end{equation}
correspondingly, we need a potential $\hat{a}$ in the Lagrangian 
\begin{equation}
    \bar{\psi}\hat{a}\gamma^3\gamma^5\psi=a_d\bar{d} \gamma^{3}\gamma^5 d+a_s\bar{s} \gamma^{3}\gamma^5 s, 
\end{equation}
considering that we only study $\phi$ and $K^{*0}$, thus, $\psi=(d,s)^{T}$, and spin potential $\hat{a}$ is also a matrix in flavor space:
\begin{equation}
    \hat{a}=
    \begin{pmatrix}
    a_d & 0 \\
    0 & a_s
    \end{pmatrix},
\end{equation}
then the spin correlation between $d$ and $\bar{s}$ is:
\begin{equation}
   \llangle \delta N_{d}\delta N_{\bar{s}} \rrangle=-VT \frac{\partial^2 }{\partial a_d\partial a_s}\Omega(a)|_{a_d,a_s\rightarrow 0},
\end{equation}
and the thermodynamic potential receives contribution from FIG.\ref{fig:omegaFD}(a), but the loops are $d$ and $s$ quark, and the only difference is the quark mass. Then we obtain
\begin{equation}
\llangle \delta N_{d}\delta N_{\bar{s}} \rrangle =N_c^2G_A^{mix}VTL(m_d)L(m_s),
\end{equation}
which is non-zero and could be either positive or negative, depending on the sign of $G_A^{mix}$. 

Above we consider  how spin polarization of quarks induce spin alignment of vector mesons, and for magnetic moment polarization together with the tensor interaction of quarks, the calculation is totally same, and here we don't repeat.

\section{The axial-vector and tensor interactions}
The interactions between quarks are complicated at finite temperature and density, especially in the presence with external fields like magnetic field or rotation， where the scalar interaction channel is mostly considered because of its close relation with chiral symmetry breaking and restoration \cite{Klevansky:1992qe,Hatsuda:1994pi}. However, all the interaction channels are supposed to exist in general and we have no conceivable reasons to ignore them. Exploring the existence of the axial-vector interaction $G_A (\Bar{\psi}\gamma^{\mu}\gamma^{5}\psi)^2$  and tensor interaction  $G_T (\Bar{\psi}i\gamma^1\gamma^2\psi)^2$, rather than relying solely on assumptions, could contribute to a clearer understanding of the physical picture. In this regard, we will now consider two potential candidates: the One-Gluon-Exchange and instanton-induced interactions.

\subsection{One-Gluon-Exchange}
The NJL-type interactions can be extracted from One-Gluon-Exchange process in QCD. Considering the non-zero gluon condensates $\langle A^2 \rangle \neq 0$ thus an effective large gluon mass $M_{g}$ arises, and the gluon propagator in Feynman gauge has form of:
\begin{equation}
    G^0_{\mu\nu}=\frac{-i g_{\mu\nu}}{q^2-M_g^2},
\end{equation}
and the color indices are ignored. However, the gluon propagator could be quite complicated, and here we consider the contribution from one-quark-loop, which leads to
\begin{equation}
    G_{\mu\nu}=G^{0}_{\mu\nu}+g^2G^{0}_{\mu\rho}\Pi^{\rho\lambda} G^{0}_{\lambda\nu}+...=\frac{-i g_{\mu\nu}}{q^2-M_g^2+\Pi},
\end{equation}
and $\Pi$ is quark loop polarization function, which simply affects the effective gluon mass:
\begin{equation}
    M_g^2\rightarrow \Tilde{M}_g^2=M_g^2-\Pi,
\end{equation}
and $\Pi$ is supposed to be functions of external conditions, for example, temperature, density, magnetic field or rotation and so on, on the other hand, $\Pi$ is also a function of $\{\mu,\nu\}$, i.e., the effective gluon mass $\Tilde{M}_g^2$ is changed for different channel.

For low energy process $q^2\ll \Tilde{M}_g^2$ where low-momentum gluon is interchanged between quarks, the four-fermion current-current interaction can be obtained by replacing the gluon propagator $G_{\mu\nu}$ by a constant: 
\begin{equation}
    J^{\mu}G_{\mu\nu}J^{\nu}\rightarrow GJ^{\mu}g_{\mu\nu}J^{\nu}=G(\bar{\psi}\gamma^{\mu}\psi)^2,
\end{equation}
where the coupling constant $G \sim 1/\Tilde{M}_g^2$. Ignore the quark loop contribution for now and $G \sim 1/M_g^2$ is a constant, then after Fierz transformation in the Dirac space, we know that scalar as well as the axial-vector interaction channels always show up \cite{Klevansky:1992qe,Liao:2012uj}:
\begin{equation}
    \mathcal{F}(\gamma^{\mu}\times\gamma_{\mu})=\mathbbm{1}\times\mathbbm{1}-\frac{1}{2}\gamma^{\mu}\times\gamma_{\mu}-\frac{1}{2}\gamma^{\mu}\gamma^{5}\times\gamma_{\mu}\gamma^{5}+\gamma^{5}\times\gamma^{5},
    \label{eq:fierzV}
\end{equation}
and the coupling constant of axial-vector interaction is half that of the scalar channel, indicating that they are of similar order. On the other hand, there is no evidence of the presence of the tensor interaction from OGE. 

The equation of Fierz transformation Eq.(\ref{eq:fierzV}) is obtained after summation over four channels $\mu=0,...,3$ and under the assumption that the coupling constants are same for each channel. Now let's perform Fierz transformation for each channel of vector interaction:
\begin{equation}
    8\mathcal{F}(\gamma^{\rho}\times \gamma^{\rho'})= g^{\rho\rho'} \sigma^{\mu\nu}\times\sigma_{\mu\nu}-4 \sigma^{\rho\mu}\times\sigma^{\rho'}_{\mu} + ..., \quad \rho=\rho',
\end{equation}
here we only show the tensor interaction terms. And it is found that the coupling constant of tensor interaction, for example, $\sim \sigma^{12}\sigma_{12}$, is 
\begin{equation}
    G_T = \frac{1}{8}(\sum_{\mu}G^{\mu}-4G^1),
\end{equation}
and in the vacuum, all $G^{\mu}$ are equal to each other then the tensor channels are canceled with each other and this is what we found in Eq.(\ref{eq:fierzV}), however, the coupling constants for $\gamma^{0}\times \gamma^{0}$ and $\gamma^{i}\times \gamma^{i}$ could be different at finite temperature and baryon chemical potential in medium, as a result, the tensor interaction will survive. 

On the other hand, the axial-vector interaction channel also receives modification from the medium, and a similar analysis leads to the coupling constant of axial-vector interaction $\gamma^3\gamma^5\times\gamma^3\gamma^5$: 
\begin{equation}
    G_A=\frac{1}{4}(G^0+G^3),
\end{equation}
and compared to the case of zero baryon chemical potential $\mu_B=0$, quark loop contribution decreases by finite $\mu_B$, thus, we make a conclusion that as collision energy decreases in heavy ion collision, where the baryon chemical potential increases, $G_T$ increases from 0 while $G_A$ decreases from its value at vacuum. On the other hand, considering that the axial-vector interaction leads to the $\rho_{00}<1/3$ while tensor interaction to $\rho_{00}>1/3$, above behavior of coupling constants could result in that $\rho_{00}$ of $\phi$ meson increases from less than $1/3$ to greater than $1/3$ as collision energy decreasing, which qualitatively explain the experiment data, and in next section we will make numerical calculation.

\subsection{Instanon-induced interactions}
The effective interactions induced by OGE does not mix flavors, which will not contribute to the spin alignment of $K^{*}$. 
On the other hand, the effective interactions of quarks can also be induced by the instantons\cite{Yu:2014sla,Schafer:1994nv}, and at different temperature, the ensembles of instantons thus the interactions of quarks are also different. What more important is instanton-induced interactions can be flavor-mixing, which could induce non-zero correlation between $s$ quark and $d$ quark thus result in spin alignment of $K^{*0}$.

At low temperature, instantons and anti-instanton are considered as non-interacting gas, and the so-called 't Hooft terms can be induced, which is important to remove the $U(1)_A$ symmetry\cite{tHooft:1976rip}. Besides, there are more terms of form\cite{Schafer:1996wv}:
\begin{equation}
    \epsilon_{f_1f_2f_3}\epsilon_{g_1g_2g_3}(\psi^{\dagger}_{f_1}\gamma_{\pm}\psi_{g_1})(\psi^{\dagger}_{f_2}\gamma_{\pm}\sigma^{\mu\nu}\psi_{g_2})(\psi^{\dagger}_{f_3}\gamma_{\pm}\sigma^{\mu\nu}\psi_{g_3}),
\end{equation}
where three flavors are considered $f_i,g_i=u,d,s$, $\gamma_{\pm}=(1\pm \gamma_5)/2$, which is the flavor-mixing tensor interaction, and in this case, the axial-vector interaction does not arise.

As temperature increases and approaches chiral phase transition temperature $T \simeq T_c$, which happens in heavy ion collision, the (anti-) instantons goes from random phase to a strongly correlated "molecular" phase ($I\bar{I}$)\cite{Ilgenfritz:1988dh,Schafer:1994nv}, and in this case, two kind of effective four-fermion axial-vector interactions can always be generated even in 3-flavor case\cite{Rapp:1999qa} :
\begin{equation}
   (\bar{\psi} \gamma^{\mu}\gamma^5\psi)^2, \quad (\bar{\psi} \gamma^{\mu}\gamma^5 \lambda^a\psi)^2,
\end{equation}
with $\lambda^a$ the Gell-Mann matrices, then we obtain the $s$ quark and $d$ quark mixing term from the second one:
\begin{equation}
   (\bar{s} \gamma^{\mu}\gamma^5 s)(\bar{d} \gamma_{\mu}\gamma^5 d), 
\end{equation}
just as we expect.

\section{Numerical estimation}
To show whether the fluctuation of spin density could induce spin alignment that comparable to the experiment measurement and help to explain the data, in this section we will perform numerical calculation and estimate its magnitude. The temperature and baryon chemical potential as functions of collision energy can be extracted from heavy ion collisions \cite{Luo:2017faz}: $\mu_B = 1.477/(1+0.343\sqrt{s})$ and $T=0.158-0.14\mu_B^2-0.04\mu_B^4$. And it is found that in collision energy region $7.7 - 200\text{GeV}$, the temperature varies from 0.134 GeV to 0.158GeV, which can be regarded as a constant, while baryon chemical potential varies from near zero to $0.4$GeV.

As we have discussed about the behavior of coupling constants of tensor and axial-vector interactions from OGE, here we simply make the assumptions:
\begin{equation}
    \label{eq:couplings}
    G_T= c \mu_B, ~ G_A=G_A^0-2c\mu_{B},
\end{equation}
which can be regarded as a Taylor expansion of $G_T/G_A$ over small $\mu_B$, and $c$ is a free parameter. Besides, non-zero baryon chemical potential would have effect on the spin alignment, however here we ignore it as we didn't consider it in above argument. On the other hand, the effective quark mass are evaluated in a 3-flavor NJL model\cite{Li:2018ygx}, and it is found that the effective mass of $s$ is $m_{s}\approx  0.425\text{GeV}$, and $m_{d} \approx 0.184\text{GeV}$ for $d$ quark. Notice that the effect from both interactions are small, here the spin alignment of $\phi$ meson is considered as: 
\begin{equation}
    \rho_{00}=\frac{1}{3}+\delta \rho_{00,A}+\delta \rho_{00,T}.
\end{equation}

\begin{figure}
 \centering
	\includegraphics[width=0.5\textwidth]{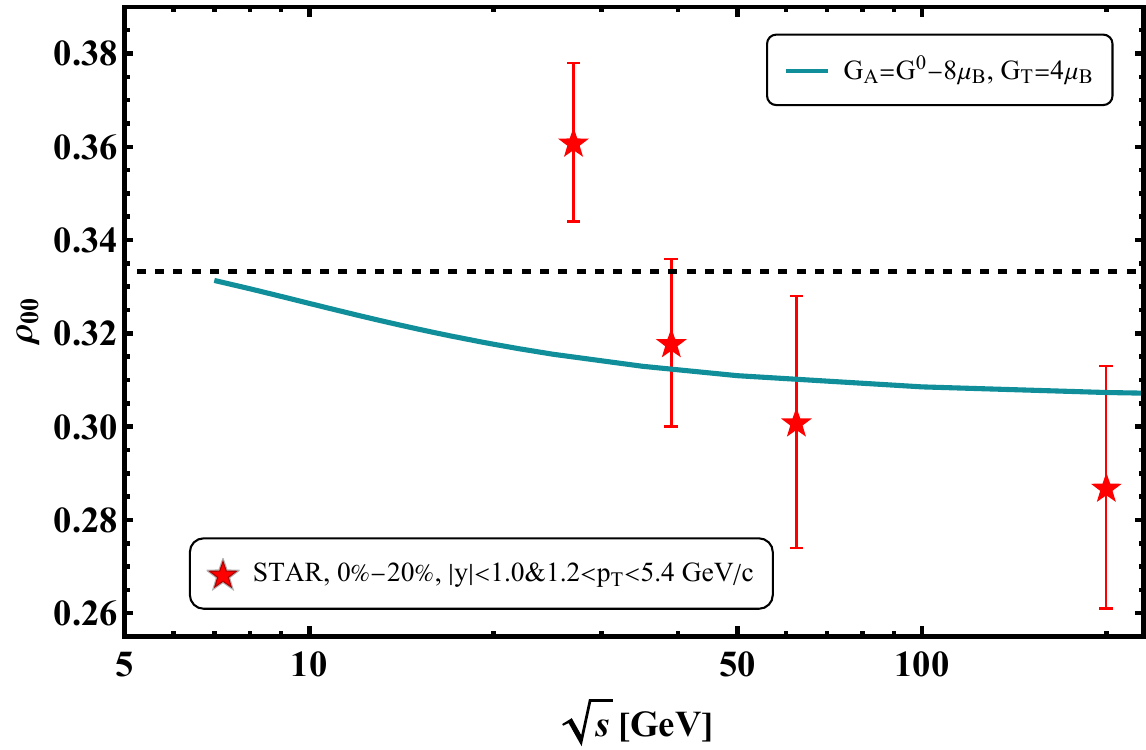}
	\caption{\justifying  Spin alignment of $\phi$ meson induced by spin density fluctuation, with both Axial-Vector and Tensor interaction.}
\label{fig:rho00}
\end{figure}

Numerical results are shown in FIG.\ref{fig:rho00}, where we take $c=4$ in Eq.(\ref{eq:couplings}), and assume the radius of QGP $r=2fm$. It is found that $\rho_{00}$ is smaller than $1/3$ significantly for $\sqrt{s}>10$GeV, and is consistent with experimental data quantitatively. On the other hand, $\rho_{00}>1/3$ can not be obtained for the current parameters, this is because that the contribution from tensor interaction is too small, while axial-vector interaction gives significant contribution. Thus, for the most central collision,  $\rho_{00}<1/3$ can be induced by the axial-vector interaction.

\section{Conclusion and Discussion}

In this work, we investigated the spin alignment of vector mesons induced by the spin density fluctuation of quarks. With the presence of axial-vector and tensor interaction of quarks, the vector meson possess spin alignment where the rotation or magnetic field are not necessary. It is found that the axial-vector(tensor) operator has a meaning of spin polarization(magnetic moment polarization), and with SP(MMP) background of quarks $a$($\kappa$), the vector mesons obtain $\rho_{00}<1/3$($\rho_{00}>1/3$), which does not depend on the sign of the background. It is worthy to point out that the influence of SP(MMP) background on the spin alignment of vector mesons is similar to rotation(magnetic field).

On the other hand,  calculating the SP background from mean field is straightforward. However, consider the presence of the axial-vector interaction of quarks, we found that the $\langle \Bar{\psi}\gamma^3\gamma^5 \psi \rangle\neq 0$ for $G_{A}\gtrsim G_S$, which means that global and homogeneous condensate of SP background can not obtained in the  mean field approximation. On the other hand, the spin alignment of vector meson induced by SP(MMP) background does not depend on the sign of the background, thus, through the fluctuation of spin density, locally non-zero SP(MMP) background can be obtained, thus $\rho_{00}$ of the vector mesons emitted from local region has deviation from $1/3$. We then calculated  $\langle (\Bar{\psi}\gamma^3\gamma^5 \psi)^2 \rangle$ instead of $\langle  \Bar{\psi}\gamma^3\gamma^5 \psi \rangle$, then non-zero spin density is obtained thus spin alignment of vector mesons appears.

We also argued the source of interactions of quarks. Generally,  all the forms of quark interaction could exist in vacuum and finite temperature, however, we still need to specify how the interaction arises. We find two possible sources, and one is from the One-Gluon-Exchange process which induces both axial-vector and tensor interactions, and both coupling constants are supposed to be functions of temperature as well as the other conditions. Another source of interaction is instanton, which could gives the tensor interaction at low temperature and axial-vector interaction around and above $T_c$. The instanton-induced interactions are also flavor-mixed, which is absent in OGE, and could explain the spin alignment of $K^{*0}$.

This mechanism could contribute to the spin alignment at central collision, as well as to the cases of non-central collision. On the other hand, the coupling constants of axial-vector and tensor interaction induced by OGE or instantons are functions of the angular velocity, magnetic field as well as the baryon chemical potential, which depends on the centrality and collision energy. 

There are still more details should be concerned. In this work, we only consider contribution of the SP potential of quarks on $\phi$ meson by changing its mass, however, the momenta are also supposed to be modified, and especially splitted in the transverse and longitudinal directions, and in this case, we can then calculate the spin alignment in the region $1\text{GeV}<p_T<5.4\text{GeV}$, more realistic results can be obtained and compared to experiments. Consider the contribution of spin density fluctuation could be important, and it is worthy to put more effort on this idea.

\textit{Acknowledgements:} We thank Xinli Sheng for helpful discussion on spin alignment of vector meson. This work is supported in part by the National Natural Science Foundation of China (NSFC) Grant Nos: 12235016, 12221005, 12147150 and the Strategic Priority Research Program of Chinese Academy of Sciences under Grant No XDB34030000.

\bibliographystyle{unsrt}
\bibliography{reference.bib}

\end{document}